%% file: MinBiasFirstPub.tex
\documentclass[preprint,12pt]{elsarticle}

\usepackage{color}

\usepackage{graphicx}
\usepackage{verbatim}
\usepackage{subfigure}
\usepackage{fancyhdr} 

\usepackage{amssymb}

\usepackage{lineno}

\usepackage{multirow}

\biboptions{square,sort&compress}

   \usepackage{epstopdf}

\usepackage{atlasphysics}
\def\nch{\ensuremath{n_{\mathrm{ch}}}}
\def\Nch{\ensuremath{N_{\mathrm{ch}}}}
\def\meanpT{\ensuremath{\langle p_\mathrm{T} \rangle}}
\def\Nsel{\ensuremath{n_{\mathrm{Sel}}}}
\def\Nselbs{\ensuremath{n_{\mathrm{Sel}}^{\mathrm{BS}}}}
\def\dzero{\ensuremath{d_\mathrm{0}}}

\def\zzsint{\ensuremath{z_\mathrm{0} \cdot \sin{\theta}}}

\journal{Phys. Lett. B}

\begin{document}

\setlength\topmargin{-1.cm}

\begin{frontmatter}

\title{Charged-particle multiplicities in $pp$ interactions\\ at $\sqrt{s} = 900$~GeV measured with the ATLAS detector at the LHC}

\author{The ATLAS Collaboration}

\newcommand{\Pythia}{PYTHIA\xspace}
\newcommand{\Phojet}{PHOJET\xspace}


\begin{abstract}
The first measurements from proton-proton collisions recorded with the ATLAS detector at the LHC are presented.
Data were collected in December 2009 using a minimum-bias trigger during collisions at a  centre-of-mass energy of 900~GeV.
The charged-particle multiplicity, its dependence on transverse momentum and pseudorapidity, and the relationship between mean transverse momentum and charged-particle multiplicity 
are measured for events with at least one charged particle in the kinematic range $|\eta|<$~2.5 and \pT~$>$~500~MeV.
The measurements are compared to Monte Carlo models of proton-proton collisions and to results from other experiments at the same centre-of-mass energy.
The charged-particle multiplicity per event and unit of pseudorapidity at $\eta =$~0 is measured to be 
$1.333\,\pm\,0.003\,$(stat.)$\,\pm\,0.040\,$(syst.), which is 5--15\% higher than the Monte Carlo models predict. 
\end{abstract}


\end{frontmatter}

\setlength\topmargin{0.cm}


\vspace{1cm}
\section{Introduction}

Inclusive charged-particle distributions have been measured in $pp$ and $p{\bar p}$ collisions at a range of different centre-of-mass energies~\cite{cmsminbias,Collaboration:2009dt,Aaltonen:2009ne,Alexopoulos:1994ag,Albajar:1989an,Abe:1989td,Ansorge:1988kn,Ansorge:1988fg,Abe:1988yu,Alner:1987wb,Ansorge:1986xq,Breakstone:1983ns,Arnison:1982ed}.  Many of these measurements have been used to constrain phenomenological models of soft-hadronic interactions and to predict properties at higher centre-of-mass energies.  
Most of the previous charged-particle multiplicity measurements were obtained by selecting data with a double-arm coincidence trigger, thus removing large fractions of diffractive events.  The 
data were then further corrected to remove the remaining single-diffractive component.  This selection is referred to as non-single-diffractive (NSD).  In some cases, 
designated as inelastic non-diffractive, the residual double-diffractive component was also subtracted.  The selection of NSD or inelastic non-diffractive charged-particle spectra 
involves model-dependent corrections for the diffractive components and for effects of the trigger selection on events with no charged particles within the acceptance of 
the detector.  The measurement presented in this paper implements a different strategy, which uses a single-arm trigger overlapping with the acceptance of the tracking volume.  Results are presented as inclusive-inelastic distributions, with minimal model-dependence, by requiring one charged particle within the acceptance of the measurement.  

This paper reports on a measurement of primary charged particles 
 with a momentum component transverse to the beam direction\footnote{The ATLAS reference system is a Cartesian right-handed co-ordinate system, with
the nominal collision point at the origin. The anti-clockwise beam direction defines the positive $z$-axis, while the positive $x$-axis is defined as pointing from the collision point to the centre of the
LHC ring and the positive $y$-axis  points upwards. The azimuthal angle $\phi$ is measured around the beam axis, and the polar angle $\theta$ is measured with respect to~the $z$-axis. The pseudorapidity
is defined as $\eta =  -\ln \tan (\theta/2) $.}
\pT~$>$~500~MeV and in the pseudorapidity range $|\eta| <$~2.5.  Primary charged particles are defined as  charged  particles with a mean lifetime $\tau  >  0.3 \times 10^{-10}$~s directly produced in $pp$ interactions or from subsequent decays of particles with  a shorter lifetime.  The distributions of  tracks reconstructed in the ATLAS inner detector were corrected to obtain the particle-level distributions:
$$
\frac{1}{N_\mathrm{ev}}\cdot  \frac{\mathrm{d} N_\mathrm{ch}}{\mathrm{d} \eta}, \ \ \ 
\frac{1}{N_\mathrm{ev}}\cdot \frac{1}{2 \pi  p_\mathrm{T}} \cdot \frac{\mathrm{d}^2 N_\mathrm{ch}}{\mathrm{d} \eta \mathrm{d} p_\mathrm{T}}, \ \ \ 
\frac{1}{N_\mathrm{ev}}\cdot \frac{\mathrm{d} N_\mathrm{ev}}{\mathrm{d} n_\mathrm{ch}} \ \ \ 
{\rm and}
\ \ \ \langle p_\mathrm{T}\rangle ~ {\mathrm vs.} ~ n_\mathrm{ch}{\rm ,}
$$
where $N_\mathrm{ev}$ is the number of events with at least one charged particle inside the selected kinematic range, \Nch\ is the total number of charged particles, \nch\ is the number of charged particles in an event  and \meanpT\ is the average \pT\ for a given number of charged particles. 
Comparisons are made to previous measurements of charged-particle multiplicities in $pp$ and $p{\bar p}$ collisions at $\sqrt{s}$~=~900~GeV centre-of-mass energies~\cite{cmsminbias,Albajar:1989an} and to Monte Carlo (MC) models.

\section{The ATLAS detector}

The ATLAS detector~\cite{:2008zzm} at the Large Hadron Collider (LHC~\cite{Evans:2008zzb})
covers  almost the whole solid angle
around the collision point with 
layers of tracking detectors, calorimeters and muon chambers. It has been designed to study 
a wide range of physics topics at LHC energies.  For the measurements presented in this
paper,  the tracking devices and the trigger system were of particular importance.

The ATLAS inner detector has full coverage in $\phi$ and covers the pseudorapidity range $|\eta|~\textless $~2.5. 
It consists of a silicon pixel detector (Pixel), a silicon microstrip detector (SCT) and a transition radiation tracker 
(TRT). These detectors cover a sensitive radial distance from the interaction point of 
50.5--150~mm, 299--560~mm and 563--1066~mm, respectively, and are immersed in a 2~T  
axial magnetic field.
The inner-detector barrel (end-cap) parts consist of 3 (2$\times$3) Pixel layers, 4 (2$\times$9)
double-layers of single-sided silicon microstrips with a 40~mrad stereo angle,
and 73 (2$\times$160) layers of
TRT straws.  
These detectors have position resolutions of typically 10, 17 and 
130~$\mu $m for the $R$-$\phi$ co-ordinate and, in case of the Pixel and SCT,  115 and 580~$\mu $m for the second measured co-ordinate.
A track from a particle traversing the barrel detector would typically have 11
silicon hits (3 pixel clusters and 8 strip clusters), and more than
30 straw hits.

The ATLAS detector has a three-level trigger system:~Level~1~(L1), Level~2 (L2) and Event~Filter~(EF). 
For this measurement, the trigger relies on the 
L1 signals from the
Beam Pickup Timing devices (BPTX) and 
the Minimum Bias Trigger Scintillators (MBTS).
The BPTX are composed of  beam pick-ups 
attached to the beam pipe $\pm$175~m from the centre of the ATLAS detector.
The MBTS are mounted at each end of the detector in front of the liquid-argon end-cap calorimeter cryostats at $z = \pm 3.56$~m and are segmented into eight sectors in azimuth 
and two rings in pseudorapidity ($2.09 < |\eta| < 2.82$ and $2.82 < |\eta| < 3.84$).   
Data were collected for this analysis using the MBTS trigger, formed from BPTX  and MBTS trigger signals.
The MBTS trigger was configured to require one hit above threshold from either side of the detector.  The efficiency of this trigger was studied with a separate prescaled L1 BPTX trigger, filtered to obtain inelastic interactions by inner detector requirements at L2 and EF.

\section{Monte Carlo simulation}
\label{mc}

Low-\pT\ scattering processes may be described by 
lowest-order perturbative Quantum Chromodynamics (QCD) two-to-two parton scatters, where the divergence of the cross section at 
\pT\ =~0  is regulated by phenomenological models.
These models include 
multiple-parton scattering, partonic-matter distributions,  
scattering between the unresolved protons and  colour reconnection~\cite{skands-2007-52}.
The PYTHIA~\cite{Sjostrand:2006za}  MC event generator 
implements several of these models.
The parameters of these models have been tuned to 
describe  charged-hadron production  and the underlying event in $pp$ and $p {\bar p}$ data  at centre-of-mass energies between 200~GeV and 1.96~TeV. 

Samples of ten million MC events were produced for single-diffractive, double-diffractive and non-diffractive processes using the PYTHIA 6.4.21 generator.
A specific set of optimised parameters, 
the  ATLAS MC09 PYTHIA tune~\cite{atlasmc09}, which employs
the MRST LO* parton density functions~\cite{Sherstnev:2007nd} and 
the \pT-ordered parton shower, 
is the reference tune throughout this paper.  These parameters were derived by tuning to underlying event and minimum-bias 
data from Tevatron at 630~\GeV and 1.8~\TeV.
The MC samples generated with this tune were used to determine detector acceptances and efficiencies 
and to correct the data.

For the purpose of comparing the present measurement 
to different phenomenological models describing minimum-bias events, 
the following additional MC samples were generated:
the ATLAS MC09c \cite{atlasmc09}  PYTHIA tune, which is an extension of the ATLAS MC09 tune optimising the strength of the colour reconnection to  describe the \meanpT\  distributions as a function of \nch, as measured by CDF in $p{\bar p}$ collisions~\cite{Aaltonen:2009ne};
the Perugia0~\cite{Skands:2009zm} PYTHIA tune, in which the soft-QCD part is tuned using only minimum-bias data from the Tevatron and CERN $p{\bar p}$ colliders;
the DW~\cite{Albrow:2006rt} PYTHIA tune, which uses the virtuality-ordered showers and was derived to describe the CDF Run II underlying event and Drell-Yan data.
Finally, the PHOJET generator \cite{phojet}
was used as an alternative model.
It 
describes low-\pT\ physics using the two-component 
Dual Parton Model~\cite{DPM1, DPM2},
which includes
soft hadronic processes described by Pomeron exchange and semi-hard processes described by perturbative parton scattering. 
PHOJET relies on PYTHIA for the fragmentation of partons.
The versions\footnote{PHOJET 1.12 with PYTHIA 6.4.21} used for this study
  were shown to agree with previous measurements~\cite{Aaltonen:2009ne,Abe:1988yu,Albajar:1989an,Abe:1989td}.

The non-diffractive, single-diffractive and double-diffractive contributions in the generated samples were  
mixed according to the generator cross sections to  fully describe the inelastic scattering. 
All the events were processed through the ATLAS detector simulation program~\cite{bib-ATLAS-simulation}, which is based on Geant4~\cite{Agostinelli:2002hh}. They were then reconstructed and analysed by the same program chain used for the data.
Particular attention was devoted to the description in the simulation of the size and position of the collision beam spot and of the detailed detector conditions  during data taking.

\section{Event selection}
\label{eventselection}

All data recorded during the stable LHC running periods between December 6 and 15, 2009, in which  the inner detector was fully operational  and the solenoid magnet was on, were used for this analysis.
During this period the beams were colliding head-on in ATLAS. 
A total of 455,593 events were collected from colliding proton bunches
in which the MBTS trigger recorded one or more counters above threshold on either side.  In order to perform an inclusive-inelastic measurement, no further requirements beyond the MBTS trigger and inner detector information were applied in this event selection.
The integrated luminosity for the final event sample, which is given here for reference only, was estimated  
using a sample of events with energy deposits in both sides of the forward and end-cap calorimeters.
The MC-based efficiency and the PYTHIA default cross section of 52.5~mb
were then used to determine the luminosity of the data sample to be
 approximately $9~\mu$b$^{-1}$, while the maximum instantaneous luminosity was
 approximately $5\times10^{26}$~cm$^{-2}$~s$^{-1}$.  The probability of additional interactions in the same bunch crossing was estimated to be less than 0.1\%.

During this data-taking period, more than $96$\% of the Pixel detector, $99$\% of the SCT and $98$\% of the TRT were operational.
Tracks were reconstructed offline within the full acceptance range $| \eta |<$~2.5 of the inner detector~\cite{Cornelissen:2008zzc,Cornelissen:2008zz}.
Track candidates were reconstructed by requiring seven or more silicon hits in total in the Pixel and SCT, and 
then extrapolated to include measurements in the TRT.
Typically, 88\% of tracks inside the TRT acceptance ($|\eta|<2$) include a TRT extension, which 
significantly improves the momentum resolution.

This paper reports results for charged particles with \pT~$>$~500~MeV, which are less prone than lower-\pT\ particles to 
large inefficiencies and their associated systematic uncertainties resulting from interactions with material inside the tracking volume.
To reduce the contribution from background events and non-primary tracks, as well as to minimise the systematic uncertainties, the following criteria were required:
\begin{itemize}
\item the presence of a primary vertex~\cite{Piacquadio:2008zzb} 
reconstructed using at least three tracks, each with:
        \begin{itemize}
	\item \pT~$>$~150~MeV, 
	\item a transverse distance of closest approach with respect to the beam-spot position
	$| d_\mathrm{0}^\mathrm{BS} | < $ 4~mm.
         \end{itemize}
\item  at least one track with: 
         \begin{itemize}
      	\item \pT~$>$~500~MeV,
	\item a minimum of  one  Pixel and six SCT hits,
	\item transverse and longitudinal impact parameters calculated with respect to the event primary vertex 
$|d_\mathrm{0}|<$~1.5~mm and $|z_\mathrm{0}|\cdot \sin \theta < $~1.5~mm, respectively. 
         \end{itemize}
\end{itemize}
%
%
%
\begin{figure}[h!]
 \begin{center}
\includegraphics[width=0.45\textwidth]{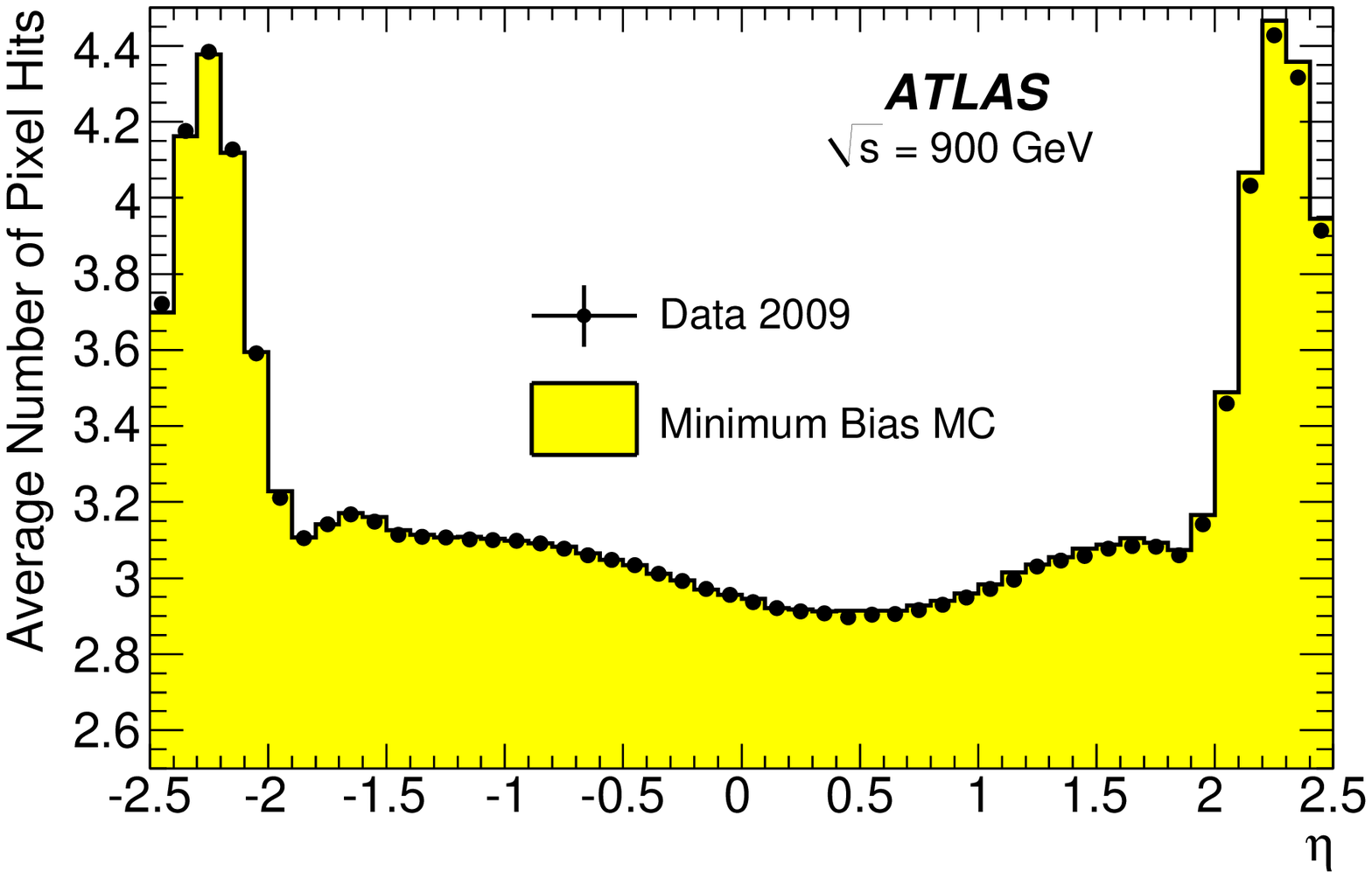}
\includegraphics[width=0.45\textwidth]{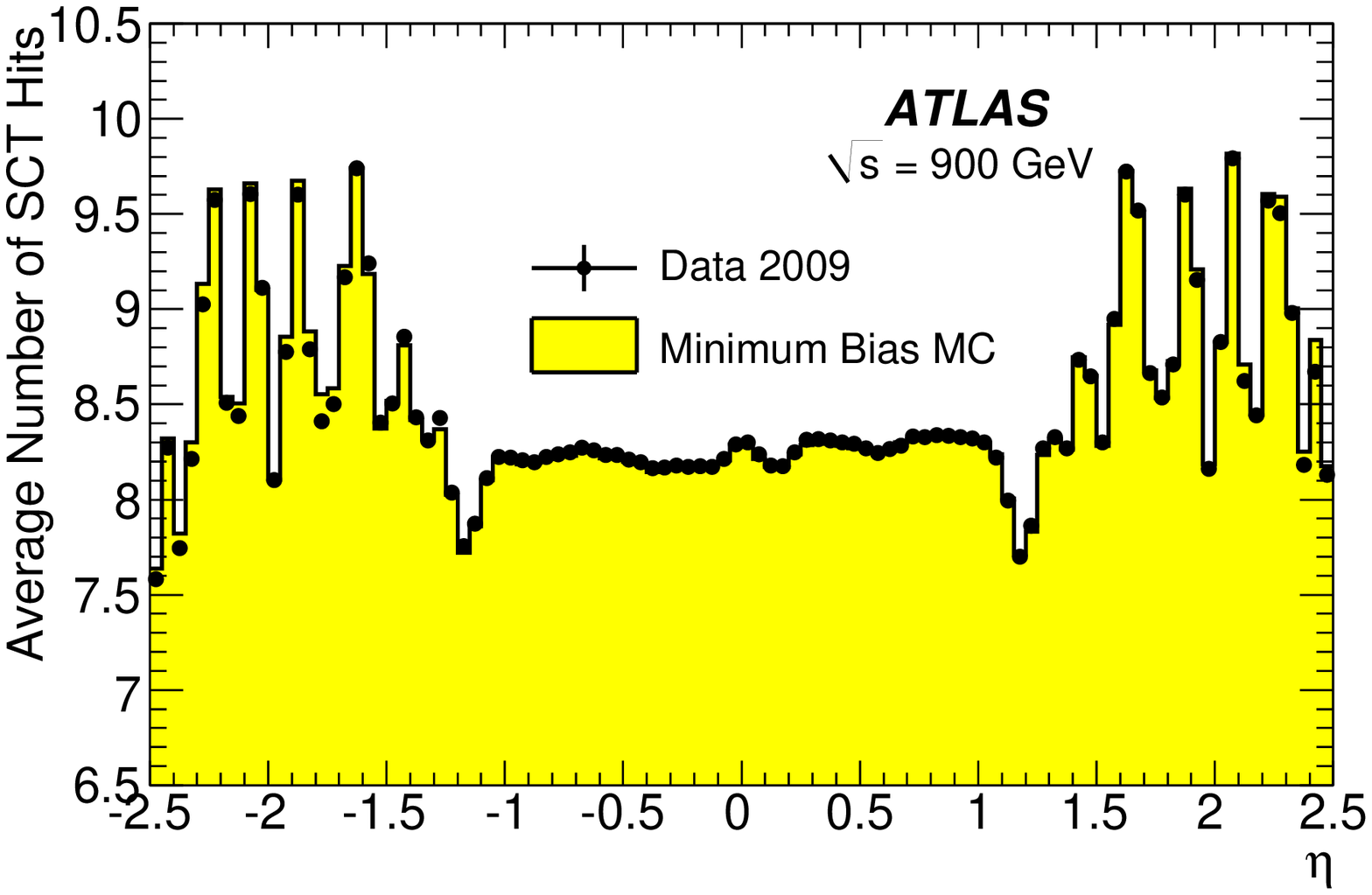}
\includegraphics[width=0.45\textwidth]{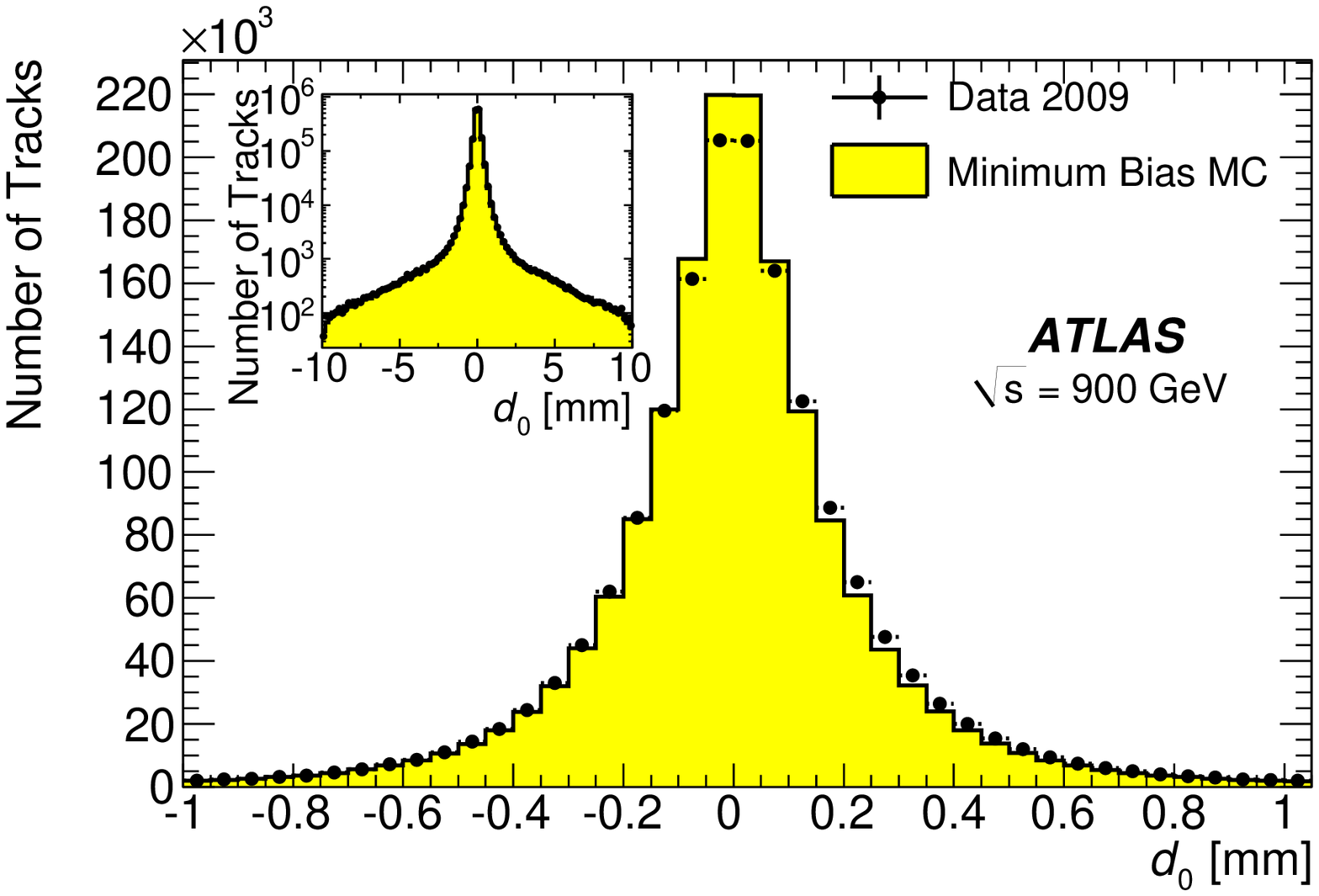}
\includegraphics[width=0.45\textwidth]{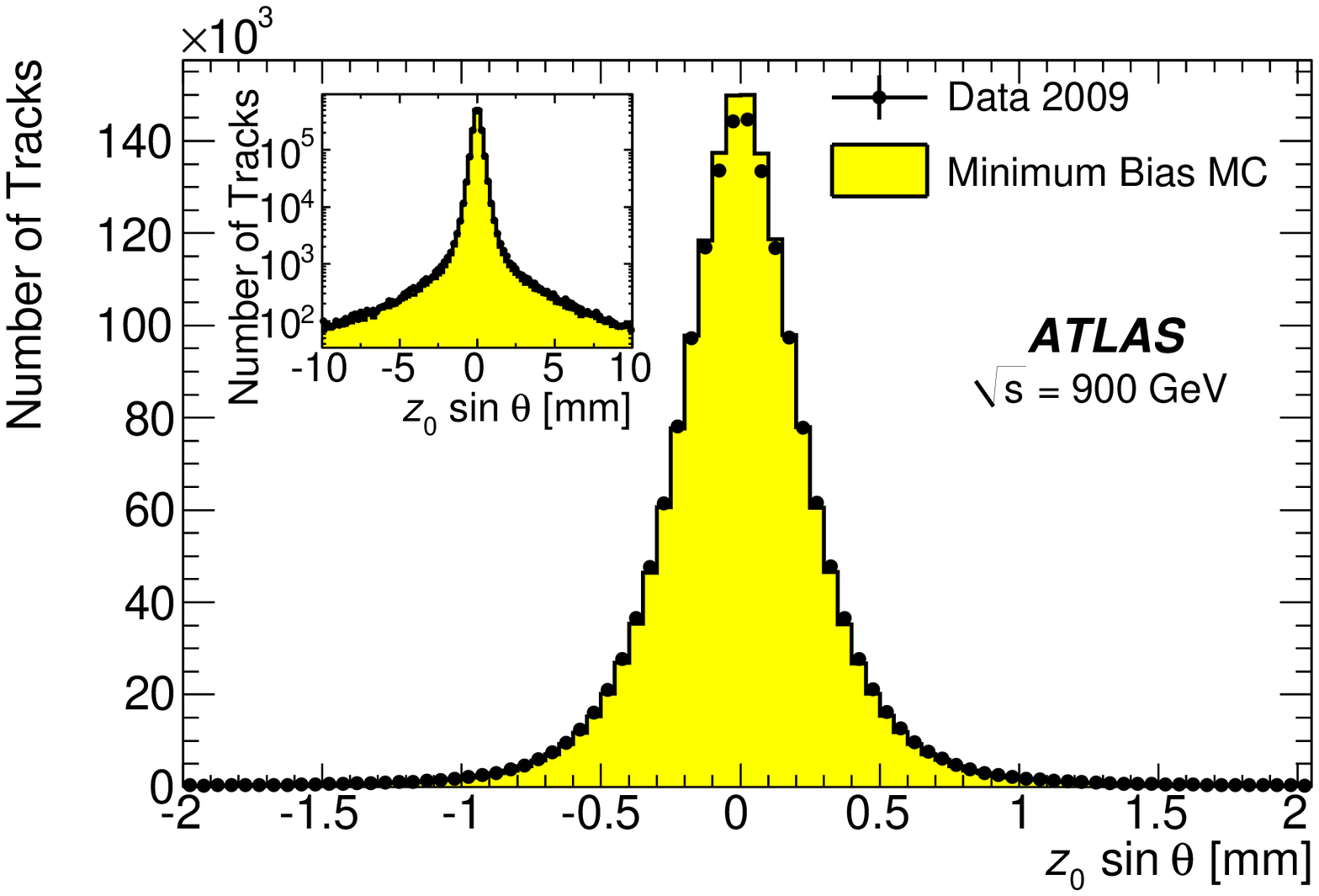}
 \end{center}
 \begin{picture} (0.,0.)
    \setlength{\unitlength}{1.0cm}
    \put ( 5.8,6.35){a)}
    \put (12.1,6.35){b)}
    \put ( 5.8,2.1){c)}
    \put (12.1,2.1){d)}
 \end{picture}
 \vspace{-1cm}
\caption{Comparison between data (dots) and minimum-bias ATLAS MC09 simulation (histograms) for the average number of Pixel hits (a) and SCT hits (b) per track as a function of $\eta$, and the transverse (c) and longitudinal (d) impact parameter distributions of the reconstructed tracks.
The MC distributions in (c) and (d) are normalised to the number of tracks in the data.  The inserts in the lower panels show the   distributions in logarithmic scale.}
\label{figure:id-quality}
\end{figure}
These latter tracks were used to produce the corrected distributions and will be referred to as selected tracks. The multiplicity of selected tracks within an event is denoted by \Nsel.~In total 326,201 events were kept after this offline selection, which contained 1,863,622 selected tracks.   The inner detector performance is illustrated in Fig.~\ref{figure:id-quality} using selected tracks and their MC simulation.  
The shapes from overlapping Pixel and SCT modules in the forward region and the inefficiency from a small number of disabled Pixel modules in the central region are well modelled by the simulation.
The simulated impact-parameter distributions describe the data to better than 10\%, including their tails as shown in the inserts of Fig.~\ref{figure:id-quality}c and d.
The difference between data and MC observed in the central region of the $d_0$ distribution is due to small residual misalignments not simulated in the MC, which are found to be unimportant for this analysis. 

Trigger and vertex-reconstruction efficiencies were  parameterized as a function of the number of tracks passing all of the track selection requirements except for the constraints with respect to the primary vertex.  Instead, the transverse impact parameter with respect to the beam spot was required to be less than 4~mm, which is the same requirement as that used in the primary vertex reconstruction preselection.  The multiplicity of these tracks in an event is denoted by  \Nselbs.

\section{Background contribution}
\label{background}

There are two possible sources of background events that can contaminate the selected sample:  cosmic rays and beam-induced background.
A limit on the fraction of cosmic-ray events recorded by the L1 MBTS trigger during data taking was determined from cosmic-ray studies, the maximum number of proton bunches, and the central trigger processor clock width of 25~ns, and was found to be smaller than $10^{-6}$.
Beam-induced background events can be produced by proton collisions with upstream collimators or with residual particles inside the beam pipe.  The L1 MBTS trigger was used to select beam-induced background events from un-paired proton bunch-crossings.  By applying the analysis selection criteria to these events,
an upper limit of $10^{-4}$ was determined for the fraction of beam-induced background events within the selected sample. 
The requirement of a reconstructed primary vertex is particularly useful to suppress the beam-induced background.

Primary charged-particle multiplicities are measured from selected-track distributions after correcting for the fraction of secondary particles in the sample. The potential background from fake tracks is found to be less than 0.1\% from simulation studies.
Non-primary tracks are mostly due to hadronic interactions, photon conversions and decays of long-lived particles.
Their contribution was  estimated using the MC prediction of the shape of the \dzero\ distribution, after normalising to data for 
tracks within 2~mm~$< |$\dzero$| <$~10~mm, i.e.~outside the range used for selecting tracks.
The simulation was found to reproduce the tails of the \dzero\ distribution of the data, as shown in Fig.~\ref{figure:id-quality}c, and the normalisation factor between data and MC was measured to be $1.00\,\pm \,$0.02$\,$(stat.)$\,\pm \,$0.05$\,$(syst.).
The MC was then used to estimate the fraction of secondaries in the selected-track sample to be (2.20$\,\pm \,$0.05$\,$(stat.)$\,\pm \,$0.11$\,$(syst.))\%.
This fraction is independent of \Nsel , but shows a dependence on \pT\ and a small dependence on $\eta$. While the correction for secondaries was applied in bins of \pT , the dependence on $\eta$ was incorporated into the systematic uncertainty.

\section{Selection efficiency}
\label{eff-corr}

The data were corrected to obtain inclusive spectra for charged primary particles satisfying the event-level requirement of at least one primary charged particle within
\pT~$>$~500~MeV and $|\eta| <$~2.5.  These corrections include inefficiencies due to trigger selection, 
vertex and track reconstruction. They also account for effects due to the momentum scale and resolution, and for the residual background from secondary tracks.

\paragraph{Trigger efficiency}

\begin{figure}[h!]
\begin{center}
\includegraphics[width=0.45\textwidth]{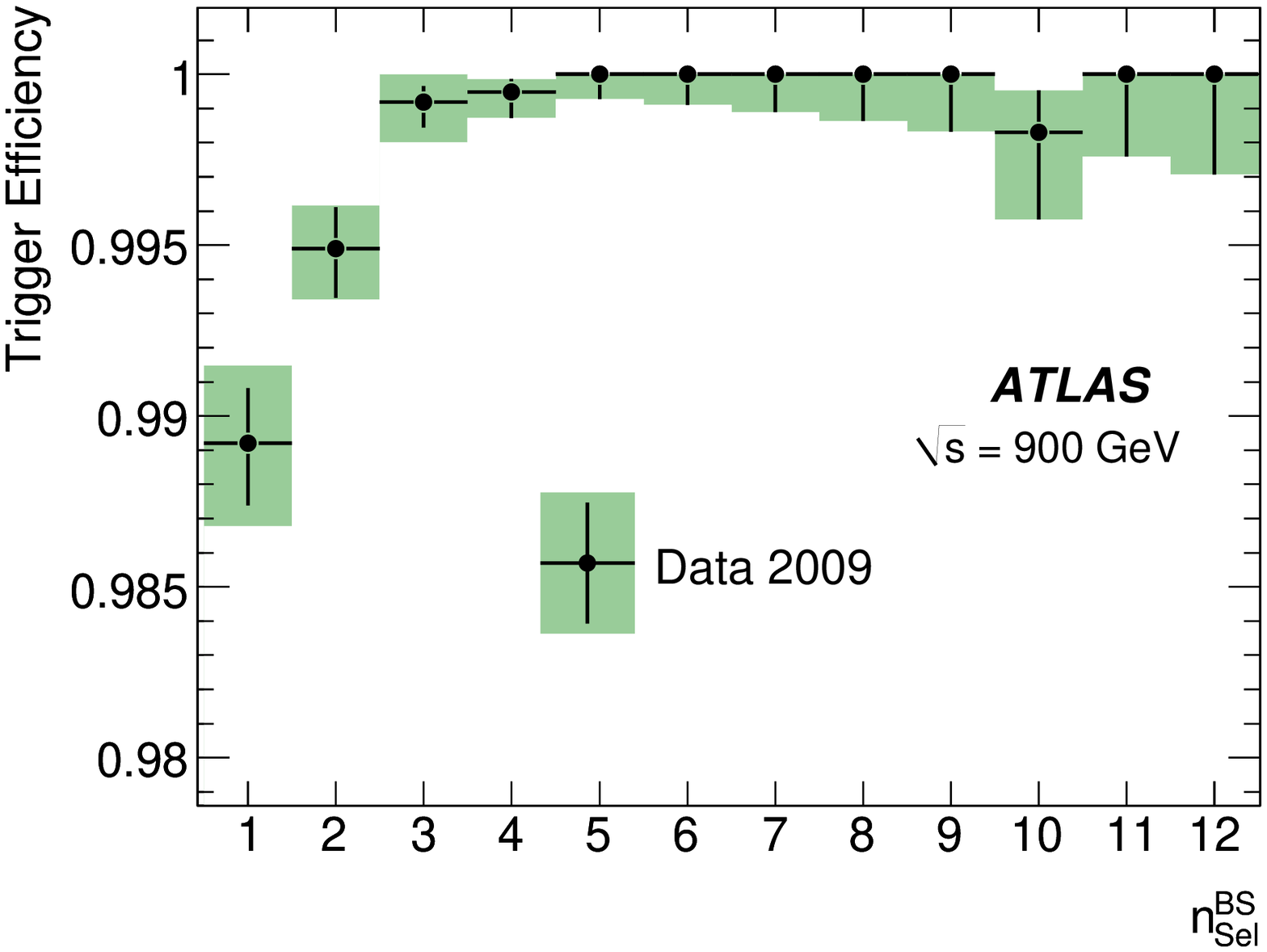}
\includegraphics[width=0.45\textwidth]{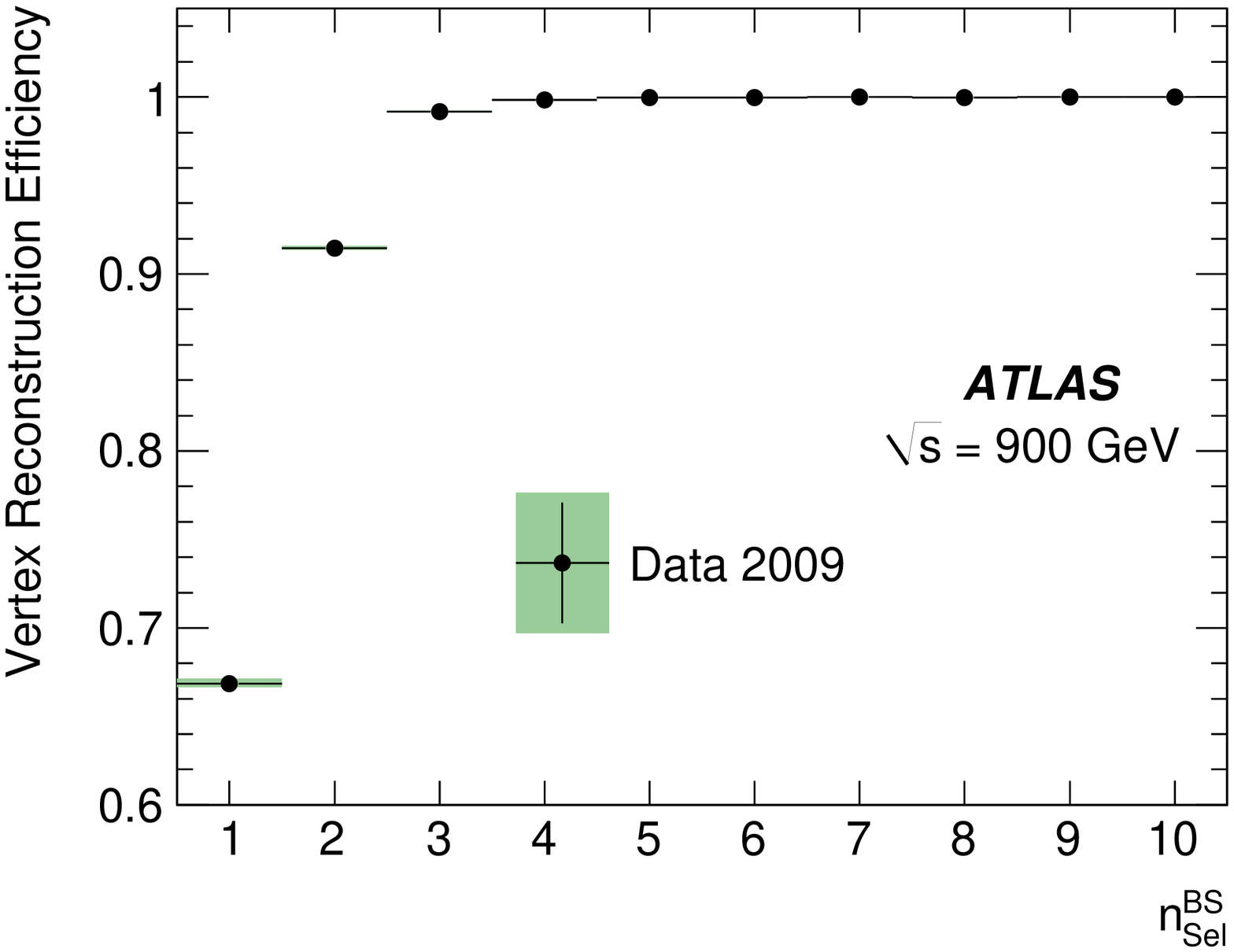}
\includegraphics[width=0.45\textwidth]{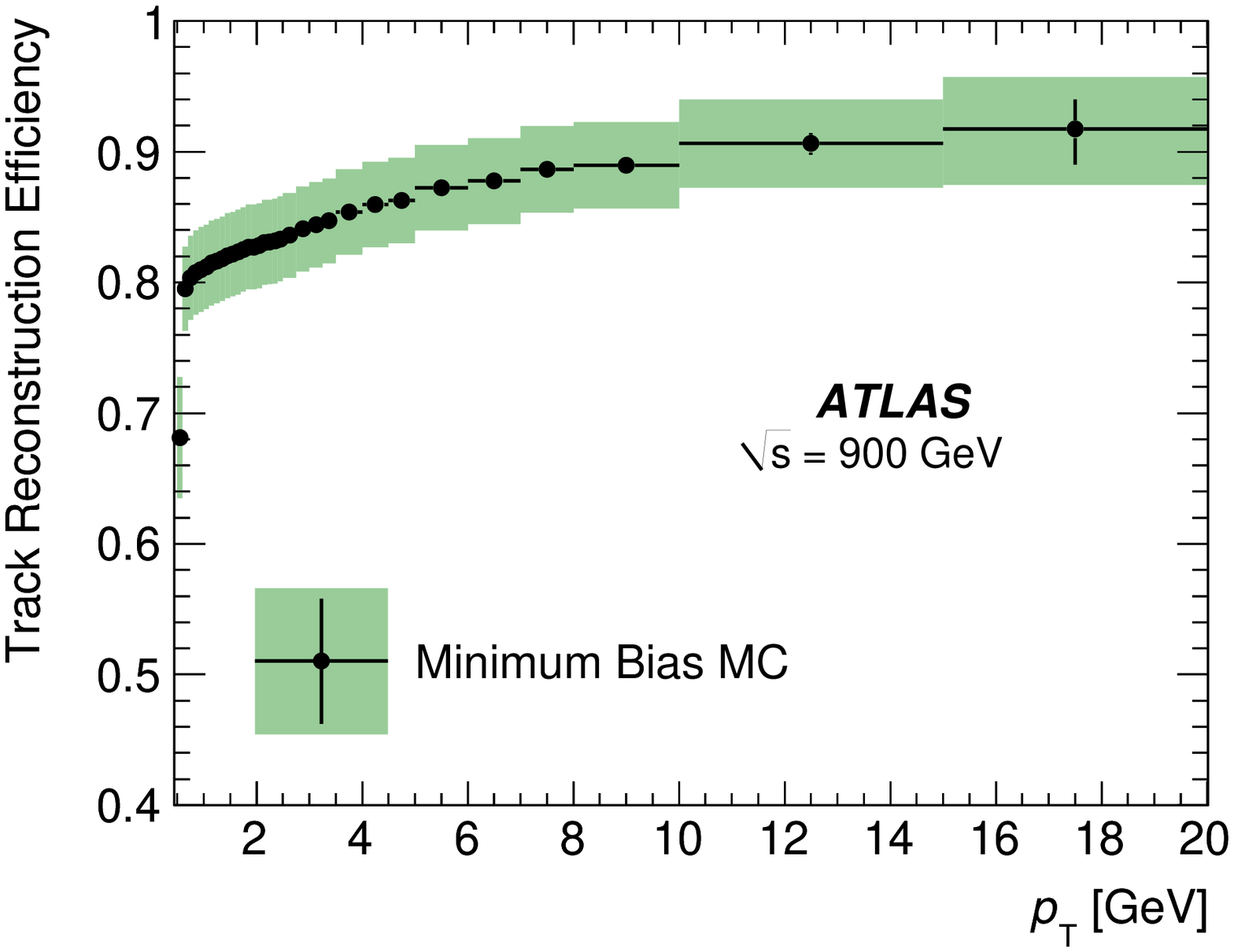}
\includegraphics[width=0.45\textwidth]{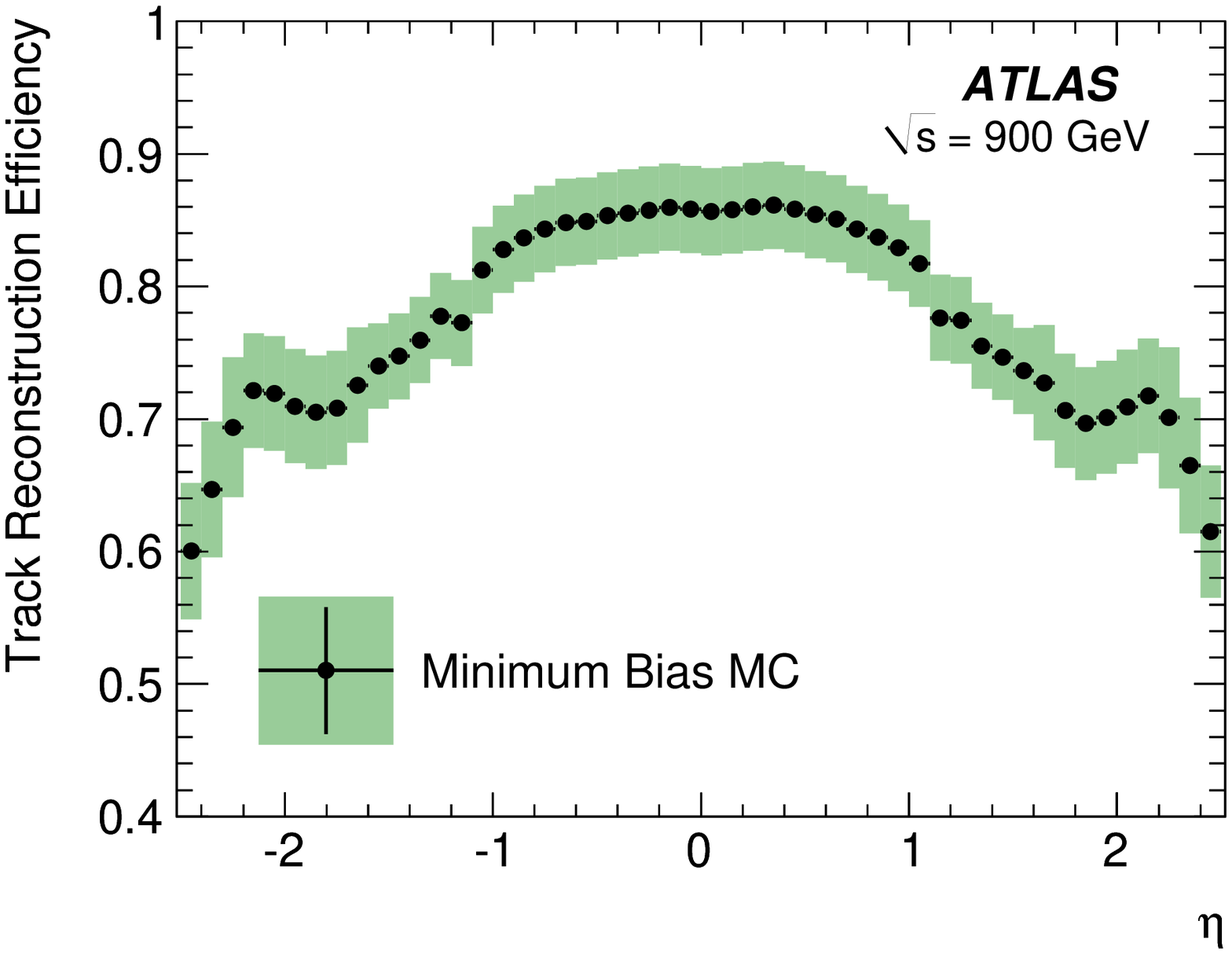}
 \end{center}
 \begin{picture} (0.,0.)
    \setlength{\unitlength}{1.0cm}
    \put ( 5.8,6.85){a)}
    \put (12.1,6.85){b)}
    \put ( 5.8,2.2){c)}
    \put (12.1,2.2){d)}
 \end{picture}
 \vspace{-1cm}
\caption{ Trigger (a) and
 vertex-reconstruction (b) efficiencies as a function of the variable \Nselbs\ defined in Section~\ref{eventselection};  track-reconstruction efficiency as a function of \pT\ (c) and of $\eta$ (d).
The vertical bars represent the statistical uncertainty, while the shaded areas represent the statistical and systematic uncertainties added in quadrature.
The two bottom panels were derived from the PYTHIA ATLAS MC09 sample.
}
 \label{figure:efficiency}
 \end{figure}

The trigger efficiency was measured from an independent data sample selected using the control trigger introduced in Section~2.  This control trigger required more than 6 Pixel clusters and 6 SCT hits at L2, and one or more reconstructed tracks with \pT~$>$~200~MeV at the EF.  The vertex requirement for selected tracks was removed for this study, to avoid correlations between the trigger and vertex-reconstruction efficiencies for L1 MBTS triggered events.  The trigger efficiency was determined by taking the ratio of events from the control trigger in which the L1 MBTS also accepted the event, over the total number of events in the control sample.
The result is shown in  Fig.~\ref{figure:efficiency}a as a function of \Nselbs.
The trigger efficiency is nearly 100\% everywhere and the requirement of this trigger does not affect the \pT\ and $\eta$ track distributions of the selected events.

%
%

\paragraph{Vertex-reconstruction efficiency}
 \label{vertexeff}

The vertex-reconstruction efficiency was determined from the data, by taking the ratio of triggered events with a reconstructed vertex to the total number of triggered events. 
It is shown in Fig.~\ref{figure:efficiency}b as a function of \Nselbs.
The efficiency amounts to approximately 67\% for the lowest bin and rapidly rises to 100\% with higher multiplicities.
The dependence of the vertex-reconstruction efficiency on the $\eta$ and \pT\ of the selected tracks was studied.  The $\eta$ dependence was found to be approximately flat for \Nselbs~$>$~1 and to decrease at larger $\eta$ for events with \Nselbs~$=$~1.  
This dependence was corrected for. No dependence on \pT\ was observed.

\paragraph{Track-reconstruction efficiency}
\label{trkeff}

The track-reconstruction efficiency in each bin of the \pT$\,$--$\,\eta$ acceptance  was determined from  MC.  The comparison of the MC and data distributions shown in Fig.~\ref{figure:id-quality} highlights their agreement.  The track-reconstruction efficiency was defined as:
$$
\epsilon_\mathrm{bin}(p_\mathrm{T},\eta) \ = \ \frac{N^\mathrm{matched}_\mathrm{rec}(p_\mathrm{T},\eta)}
                                   {N_\mathrm{gen}(p_\mathrm{T},\eta)},
$$
where  \pT\ and $\eta$ are generated quantities, and $N^\mathrm{matched}_\mathrm{rec}(p_\mathrm{T},\eta)$  and $N_\mathrm{gen}(p_\mathrm{T},\eta)$ are the number of reconstructed tracks in a given bin matched to a generated charged particle and the number of generated charged particles in that bin, respectively. 
The matching between a generated particle and a reconstructed track was done using a cone-matching algorithm in the $\eta\,$--$\,\phi$ plane, associating the particle to the track with the smallest $\Delta R \ = \ \sqrt {(\Delta\phi)^2+(\Delta \eta )^2}$
within a cone
of radius 0.05.
The resulting reconstruction efficiency as a function of \pT\ integrated over $\eta$ is
shown in Fig.~\ref{figure:efficiency}c. 
The drop to $\approx$~70\% for $\pT<$ 600 MeV is an artefact of the \pT\ cut at the pattern-recognition level and is discussed in Section~\ref{systematics}.
The reduced track-reconstruction efficiency in the region $|\eta| > 1$ 
(Fig.~\ref{figure:efficiency}d)
is mainly
due to the presence of more material in this region.
These inefficiencies include a 5\% loss due to the track selection used in this analysis, approximately half of which is due to the silicon-hit requirements and half to the impact-parameter requirements. 


\section{Correction procedure}

The effect of events lost due to the trigger and vertex requirements can be corrected for using an event-by-event weight:

$$
w_\mathrm{ev}(n_\mathrm{Sel}^\mathrm{BS}) =  \frac{1}{\epsilon_\mathrm{trig}(n_\mathrm{Sel}^\mathrm{BS})} \cdot \frac{1}{\epsilon_\mathrm{vtx}(n_\mathrm{Sel}^\mathrm{BS})}, 
$$
where $ \epsilon_\mathrm{trig}(n_\mathrm{Sel}^\mathrm{BS})$ and $\epsilon_\mathrm{vtx}(n_\mathrm{Sel}^\mathrm{BS})$ are the trigger and vertex reconstruction efficiencies discussed in Section~\ref{eff-corr}. 
The vertex-reconstruction efficiency  for events with \Nselbs~$= 1$ includes an $\eta$-dependent correction which was derived from the data.

The \pT\ and $\eta$ distributions of selected tracks were corrected  on a track-by-track basis using the weight:
$$
w_\mathrm{trk}(p_\mathrm{T}, \eta) =  \frac{1}{\epsilon_\mathrm{bin}(p_\mathrm{T}, \eta)}\cdot(1-f_{\rm sec}(p_\mathrm{T}))\cdot(1-f_{\rm okr}(p_\mathrm{T}, \eta)), 
$$
where $\epsilon_{bin}$
is the track-reconstruction efficiency described in Section~\ref{trkeff} and 
$f_{\rm sec}(p_\mathrm{T})$ is the fraction of secondaries determined as described in Section~\ref{background}. 
The fraction of selected tracks for which the corresponding primary particles are outside the kinematic range, $f_{\rm okr}(p_\mathrm{T}, \eta)$, originates from resolution effects and has been estimated from MC.
Bin migrations were found to be due solely to reconstructed track momentum resolution and were corrected by using the resolution function taken from MC.

In the case of the distributions versus~\nch, a track-level correction was applied by using Bayesian unfolding~\cite{D'Agostini:1994zf} to correct back to the number of charged particles. 
A matrix $M_\mathrm{ch,Sel}$, which expresses the probability that a multiplicity 
of selected tracks \Nsel\ is due to \nch\ particles, was populated using MC and 
applied to obtain the \nch\ distribution from the data. 
The resulting distribution was then used to re-populate  
the matrix and the correction was re-applied. This procedure was repeated without a regularisation term and converged after four iterations, when the change in the distribution between iterations was found to be less than 1\%. 
It should be noted that the matrix cannot correct for events which are lost due to track-reconstruction inefficiency. To correct for these missing events, a correction factor 
$1/(1-(1-\epsilon(n_\mathrm{ch}))^{n_\mathrm{ch}})$ was applied, where $\epsilon(n_\mathrm{ch})$ is the average track-reconstruction efficiency.

In the case of the \meanpT~versus~\nch\ distribution, 
each event was weighted by $w_\mathrm{ev}(n_\mathrm{Sel}^\mathrm{BS})$. 
For each $n_{\mathrm {Sel}}$ a MC-based correction was applied to convert the reconstructed average $p_\mathrm{T}$ to the average $p_\mathrm{T}$ of primary charged particles.
Then the matrix $M_\mathrm{ch,Sel}$ 
was applied as described above.


\section{Systematic uncertainties}
\label{systematics}

Numerous detailed studies have been performed to understand possible sources 
of systematic uncertainties. The main contributions are discussed
below.

\paragraph {Trigger}

The trigger selection dependence on the \pT\ and $\eta$ distributions of reconstructed tracks was found to be flat within the statistical uncertainties of the data recorded with the control trigger.  The statistical uncertainty on this result was taken as a systematic uncertainty of 0.1\% on the overall trigger efficiency.

Since there is no vertex requirement in the data sample used to measure the trigger efficiency, it is not possible to make the same impact-parameter cuts as are made on the final selected tracks. Therefore 
the trigger efficiency was measured using impact-parameter constraints with respect to the primary vertex or the beam spot and compared to that obtained without such a requirement.  The difference was taken as a systematic uncertainty of 0.1\% for $n_\mathrm{Sel}^\mathrm{BS} \leq$ 3.

The correlation of the MBTS trigger with the control trigger used to select the data sample for the trigger-efficiency determination was studied using the simulation.  The resulting systematic uncertainty was found to affect only the case \Nselbs~$=1$ and amounts to 0.2\%.

\paragraph {Vertex reconstruction}

The run-to-run variation of the vertex-reconstruction efficiency was found to be within the statistical uncertainty.  The contribution of beam-related backgrounds to the sample selected without a vertex requirement was estimated by using non-colliding bunches. It was  found to be  0.3\% for  \Nselbs~$=1$  and smaller than 0.1\% for higher multiplicities, and  was assigned as a systematic uncertainty.  This background contribution is larger than that given in Section~\ref{background}, since a reconstructed primary vertex was not required for these events.

\paragraph{Track reconstruction and selection}
Since the track-reconstruction efficiency is determined from MC, the main systematic uncertainty results from the level of disagreement between data and MC. 

Three different techniques  to associate generated particles to reconstructed tracks
were studied: a cone-matching algorithm, an evaluation of the fraction of simulated hits associated to a reconstructed track and an inclusive technique using a correction for secondary particles.  A systematic uncertainty of 0.5\% was assigned from the difference between the cone-matching and the hit-association methods.

A detailed comparison of track properties in data and simulation was performed by varying the track-selection criteria. 
The largest deviations between data and MC were observed by varying the \zzsint\ selection requirement, and by varying the constraint on the number of SCT hits. These deviations are generally smaller than 1\% and rise to 3\% at the edges of the $\eta$ range.

The systematic effects of misalignment were studied by smearing simulation samples by the expected residual misalignment and by comparing the performance of two alignment algorithms on tracks reconstructed from the data.  Under these conditions the number of reconstructed tracks was measured and the systematic uncertainty on the track reconstruction efficiency due to the residual misalignment was estimated to be less than 1\%.

To test the influence of an imperfect description of the detector material in the simulation, two additional MC samples with approximately 10\% and 20\% increase in radiation lengths of the material in the Pixel and SCT active volume were used.  
The impact of excess material in the tracking detectors was studied using the tails of the impact-parameter distribution, the length of tracks, and the change in the reconstructed \kshort\ mass as a function of the decay radius, the direction and the momentum of the \kshort . The  MC with nominal material was found to describe the data best.  The data were found to be consistent with a 10\% material increase in some regions, whereas the 20\% increase was excluded in all cases. 
The efficiency of matching full tracks to track segments reconstructed in the Pixel detector was also studied.  The comparison between data and simulation was found to have good agreement across most of the kinematic range.  Some discrepancies found for $|\eta|$~$>$~1.6 were included in the systematic uncertainties.
From all these studies a systematic uncertainty on the track reconstruction efficiency of 3.7\%, 5.5\% and 8\% was assigned to the pseudorapidity regions 
$|\eta|$~$<$~1.6, 1.6~$<$~$|\eta|$~$<$~2.3 and $|\eta|$~$>$~2.3, respectively.

The track reconstruction efficiency shown in Fig.~\ref{figure:efficiency}c rises sharply 
in the region
 $500 <$~\pT~$< 600$~MeV.  The observed turn-on curve is produced by the initial pattern recognition step of track reconstruction and its associated \pT\ resolution,
 which is considerably worse than the final \pT\ resolution.  
 The consequence is that  some particles which are simulated with \pT~$> 500$~MeV are reconstructed with momenta below the  selection requirement. This effect reduces the number of selected tracks. The shape of the threshold was studied in data and simulation and a systematic uncertainty of 5\% was assigned to the first \pT\ bin. 

In conclusion, an overall relative systematic uncertainty of 4.0\% was assigned to the track reconstruction efficiency for most of the kinematic range of this measurement,
while 8.5\% and 6.9\% were assigned to the highest $|\eta|$ and to the lowest \pT\ bins, respectively.

\paragraph {Momentum scale and resolution}

To obtain corrected distributions of charged particles, the scale and resolution uncertainties in the reconstructed \pT\ and $\eta$ of the selected tracks have to be taken into account. 
Whereas the uncertainties for the $\eta$ measurement were found to be negligible, those for the \pT\ measurement are in general more important. 
The inner detector momentum resolution was taken from MC as a function of \pT\ and $\eta$.
It was found to vary between 1.5\% and 5\% in the range relevant to this analysis.
The uncertainty was estimated by comparing with MC samples with a uniform scaling
of 10\% additional material at low \pT\ and with large misalignments at higher \pT . Studies of the
width of the mass peak for reconstructed \kshort\ candidates in the data show that  these assumptions are conservative.
The reconstructed momentum scale was checked by comparing the measured 
value of the \kshort\ mass to the MC. The systematic uncertainties from
both the momentum resolution and scale were found to have no significant effect on the final results.

\paragraph {Fraction of secondaries}

The fraction of secondaries was determined as discussed in Section~\ref{background}. The associated systematic uncertainty was estimated by varying the range 
of the impact parameter distribution that was used to normalise the MC, and by fitting separate distributions for weak decays and material interactions. 
The systematic uncertainty includes a small contribution due to the $\eta$ dependence of this correction. The total uncertainty is 0.1\%.

\paragraph {Correction procedure}

Several independent tests were made to assess the model dependence of the correction matrix $M_\mathrm{ch,Sel}$ and the resulting systematic uncertainty.
In order to determine the sensitivity to the \pT\ and $\eta$ distributions,
the matrix was re-populated using the other MC parameterizations described in Section~\ref{mc}
and by varying the track-reconstruction efficiency by $\pm$~5\%. 
The correction factor for events lost due to the track-reconstruction inefficiency was varied by the same amount and treated as fully correlated.
For the overall normalisation, this
leads to an uncertainty of 0.4\%  due to the model dependence and of 1.1\% due to
the track-reconstruction efficiency. The size of the systematic uncertainties on \nch\ increases with
the multiplicity. 
  
The correction for  the \meanpT\ was also studied using the different PYTHIA tunes and PHOJET.
The change was found to be less than 2\% over the whole sample.

As the track-reconstruction efficiency depends on the particle type, 
the uncertainty in the composition of the charged particles in the minimum-bias MC sample was studied. 
The relative yields of pions, kaons and protons in the simulation were separately varied by $\pm 10$\%.  These variations, combined with changing the  fraction of electrons and muons by a factor of three, resulted in a systematic uncertainty of 0.2\%.
\\

The systematic uncertainty on the normalisation and on the number of charged particles were treated separately.
In each of these two groups the systematic uncertainties were
 added in quadrature. 
These were then combined taking into account their anti-correlation and were propagated to the final distributions.
Table~\ref{tab:sysSummary} summarises the various contributions to the systematic uncertainties on the charged-particle density at $\eta$ = 0.

\begin{table}[h!]
 \begin{center}
   \renewcommand{\arraystretch}{1.3}
   \begin{tabular}{ | l | c | }

   \hline\hline
   \multicolumn{2}{|c|}{ {\bf Systematic uncertainty on the number of events, \boldmath{$N_\mathrm{ev}$}}}  \\
   \hline
    Trigger efficiency &$<\,0.1\%$ \\
    Vertex-reconstruction efficiency & $<\,0.1\%$ \\
    Track-reconstruction efficiency & $\,\,1.1\%$ \\
    Different MC tunes & $\,\,0.4\%$ \\
        \hline
    Total uncertainty on $N_\mathrm{ev}$ &  $\,\,1.2\%$ \\
    \hline\hline
   \multicolumn{2}{|c|}{ {\bf Systematic uncertainty on \boldmath{${(\mathbf 1/N_\mathrm{ev})\cdot}($d$N_\mathrm{ch}/$d$\eta$) at $\eta$ = 0}} }  \\
   \hline

   Track-reconstruction efficiency & $\,\,4.0\%$ \\
   Trigger and vertex efficiency & $<\,0.1\%$ \\
   Secondary fraction & $ \,0.1\%$ \\
   Total uncertainty on $N_\mathrm{ev}$   & $\,-1.2\%$ \\
   \hline
  Total uncertainty on (${1/N_\mathrm{ev})\cdot}($d$N_\mathrm{ch}/$d$\eta$) at $\eta$ = 0 &   $\,\,2.8\%$   \\
   \hline\hline
   \end{tabular}
   \caption{\label{tab:sysSummary} Summary of systematic
uncertainties on the number of events, $N_\mathrm{ev}$, and on
the charged-particle density ${(1/N_\mathrm{ev})\cdot}(\mathrm{d}N_\mathrm{ch}/\mathrm{d}\eta)$ at $\eta$~=~0.
The uncertainty on $N_\mathrm{ev}$ is  anticorrelated with $\mathrm{d}N_\mathrm{ch}/\mathrm{d}\eta$.
All other sources are assumed to be uncorrelated.}
 \end{center}
\end{table}


\section{Results}

\begin{figure}[htbp]
 \begin{center}
\includegraphics[width=0.45\textwidth]{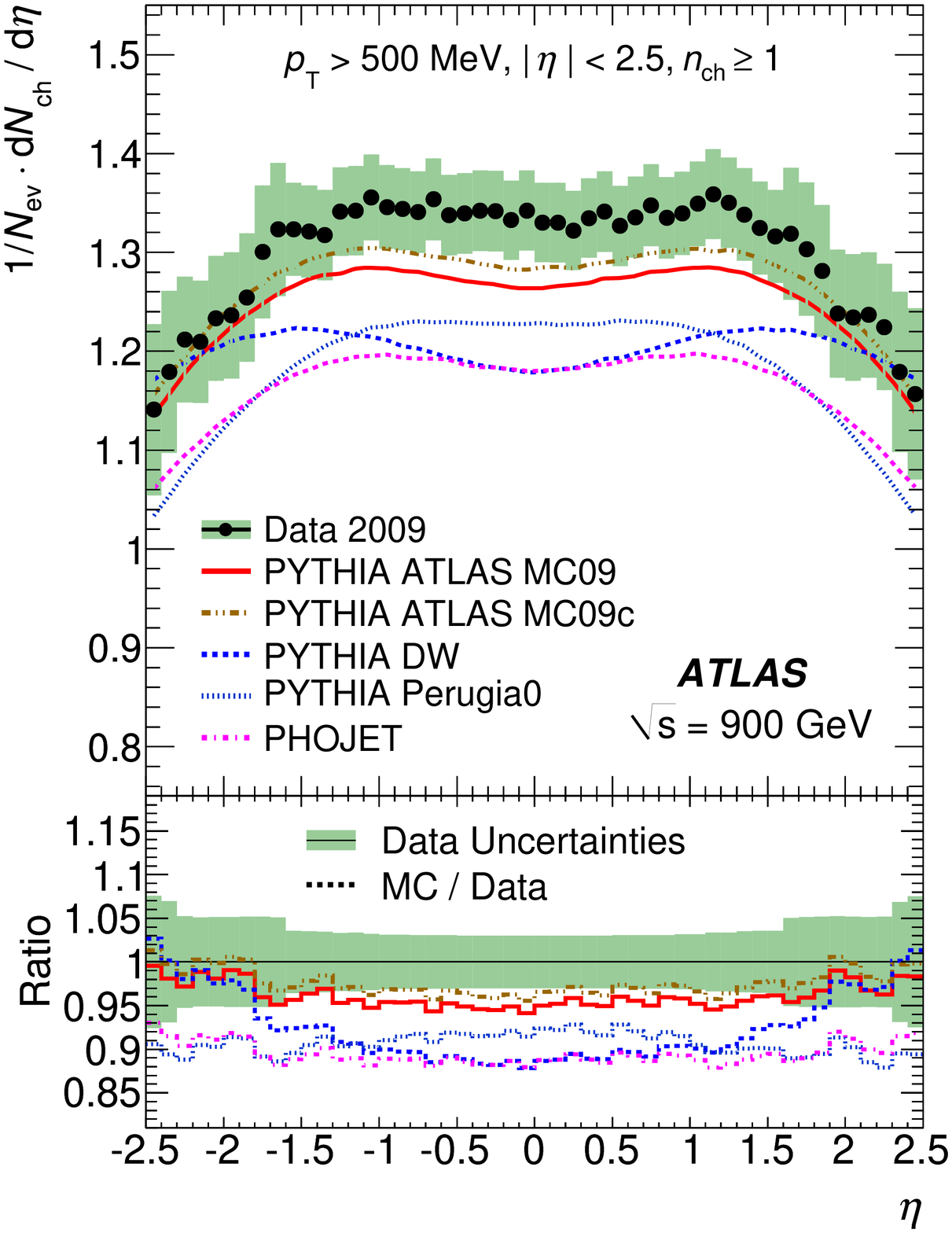}
\includegraphics[width=0.45\textwidth]{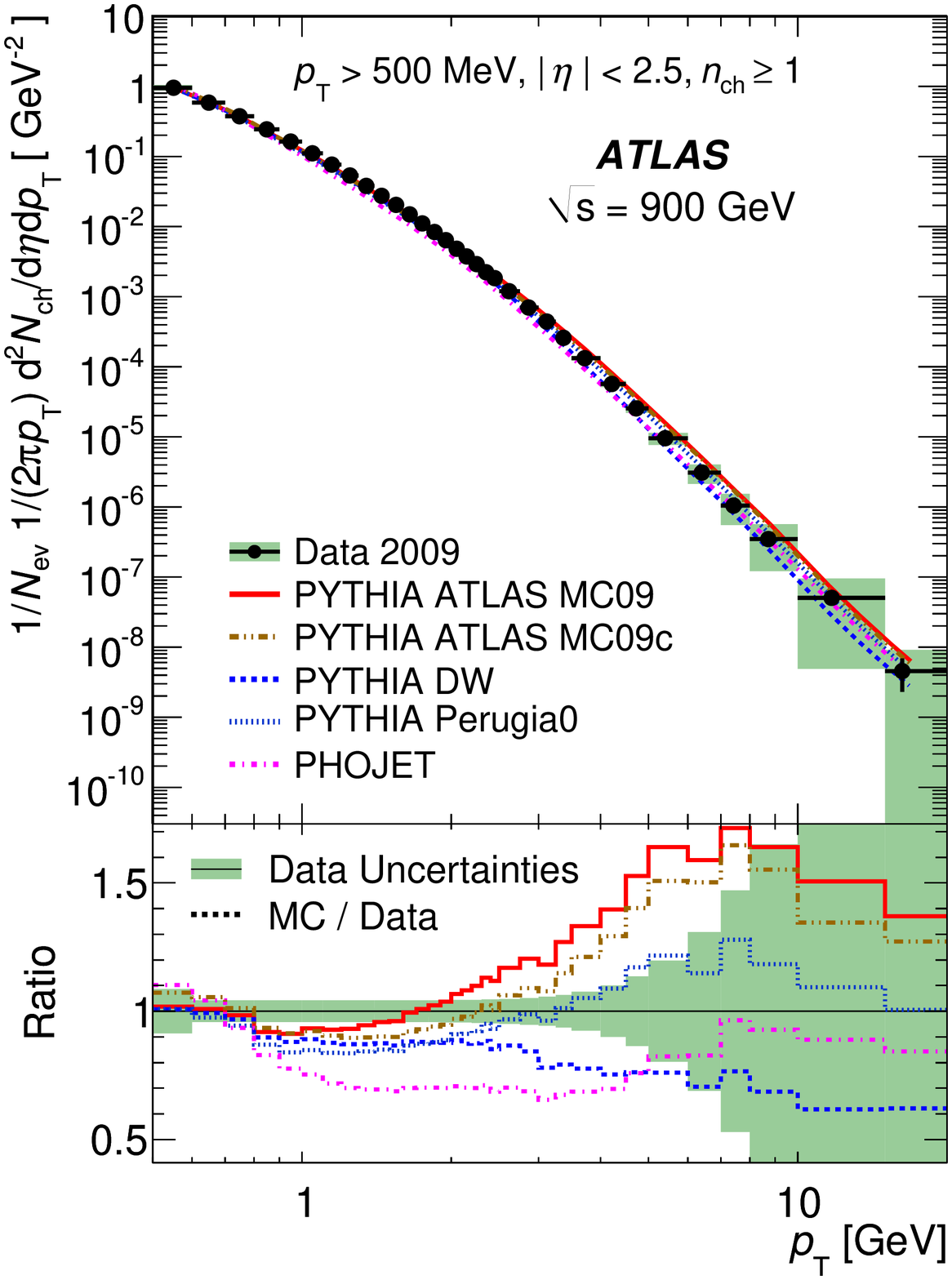}
\includegraphics[width=0.45\textwidth]{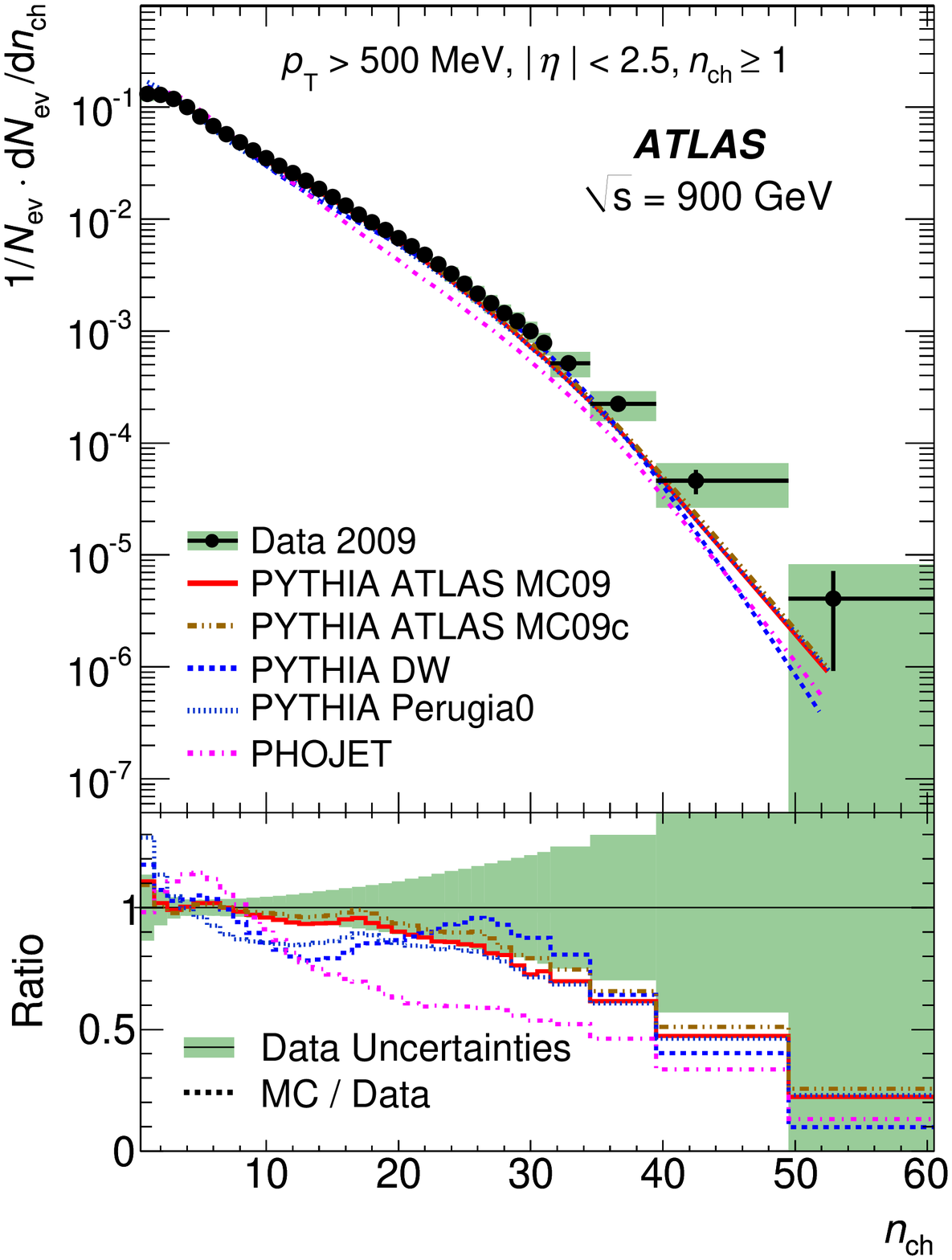}
\includegraphics[width=0.45\textwidth]{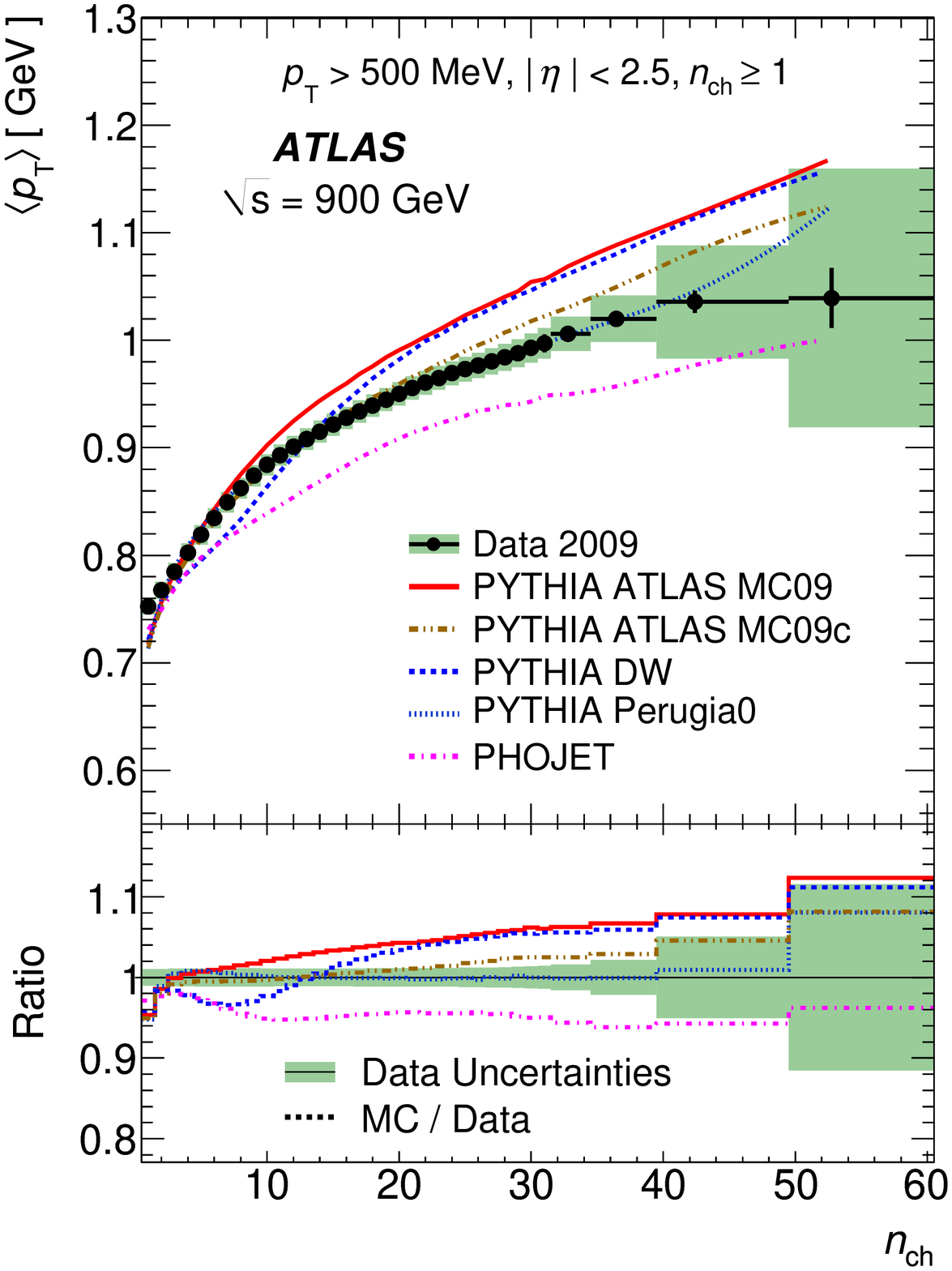}
 \end{center}
 \begin{picture} (0.,0.)
    \setlength{\unitlength}{1.0cm}
    \put ( 5.8,16.0){a)}
    \put (12.1,16.0){b)}
    \put ( 5.8,8.){c)}
    \put (12.1,8.){d)}
 \end{picture}
 \vspace{-1cm}
\caption{Charged-particle multiplicities for events with \nch~$\geq$~1 within the kinematic range \pT~$>$~500~MeV and $|\eta| < $~2.5.
The panels show the charged-particle multiplicity as a function of pseudorapidity (a) and of the  transverse momentum (b),  the charged-particle multiplicity (c), and the average transverse momentum as a function of the number of charged particles in the event (d).
The dots represent the data and the curves the predictions from different MC models. The vertical bars represent the statistical uncertainties, while the shaded areas show statistical and systematic uncertainties added in quadrature.
The values of the ratio histograms refer to the bin centroids.
}
\label{fig:results}
\end{figure}

The corrected distributions for primary charged particles for events with \nch~$\geq$ 1 in the
kinematic range \pT~$>$~500~MeV and $|\eta| < $~2.5
are shown in Fig.~\ref{fig:results},
where they are compared to predictions of models tuned to a wide range of measurements. 
The data are presented as inclusive-inelastic distributions with minimal model-dependent
corrections to facilitate the comparison with models. 

The  charged-particle pseudorapidity density 
is shown in Fig.~\ref{fig:results}a.
It is measured to be approximately flat in the range $|\eta| <$~1.5,
with an average value of 
$1.333\,\pm\,0.003\,$(stat.)$\,\pm\,0.040\,$(syst.)~charged particles per event and unit of pseudorapidity in the range $|\eta|<$~0.2. 
The particle density is found to drop at higher values of $|\eta|$.  All MC tunes discussed in this paper are lower than the data by 5--15\%, corresponding to approximately 1--4 standard deviations.  The shapes of the models are approximately consistent with the data with the exception of PYTHIA DW.

%
%


The \Nch\ distribution in bins of \pT\ is shown in Fig.~\ref{fig:results}b 
and is constructed by weighting each entry by $1/\pt$.  
The MC models do not reproduce the data
 for \pT~$>0.7$~GeV.
The most significant difference is seen for the PHOJET generator.

The multiplicity distribution
as a function of \nch\ is shown in Fig.~\ref{fig:results}c. 
The PYTHIA models show an excess of events with \nch~$=1$ with respect to the data, while the fraction of events with \nch~$\gtrsim$~10 is consistently lower than the data.  The net effect is that the integral number of charged particles predicted by the models are below that of the data (Fig.~\ref{fig:results}a and \ref{fig:results}b).  The PHOJET generator successfully models the number of events with \nch~$=1$, while it deviates from the data distributions at higher values of \nch.

The average \pT\ as a function of \nch\  is illustrated in Fig.~\ref{fig:results}d. 
It 
is found to increase with increasing \nch\ 
and a change of slope is observed around \nch~$=$~10.
This behaviour was already observed by the CDF experiment in $p {\bar p}$ collisions at 1.96~TeV~\cite{Aaltonen:2009ne}. 
The Perugia0 parameterization, which was tuned using CDF minimum-bias data at 1.96~TeV, describes the data well.
The other models fail to describe the data below \nch~$\approx$~25, with the exception of the PYTHIA-MC09c tune.

\begin{figure}[htbp]
\begin{center}
\resizebox{0.9\textwidth}{!}{\includegraphics{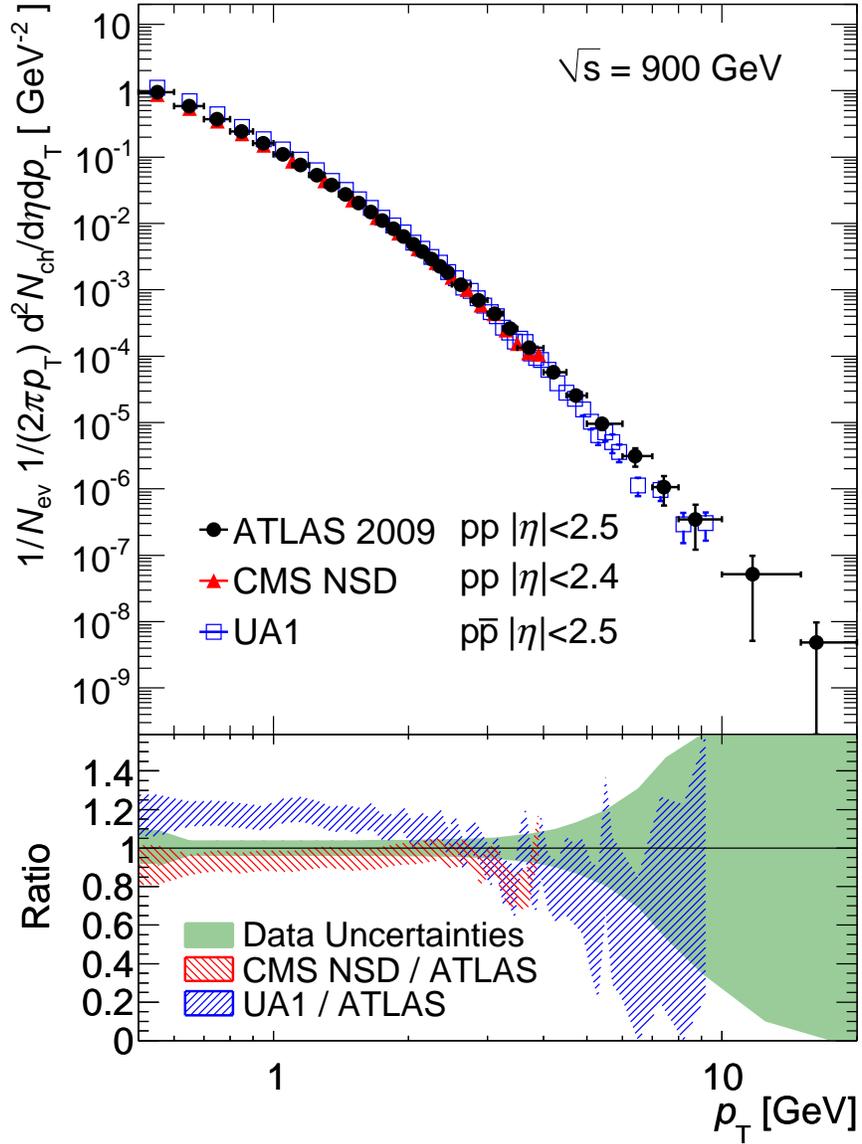}}
\caption{The measured \pT\ spectrum of charged-particle multiplicities. The ATLAS $pp$ data (black dots) are compared to the UA1 $p{\bar p}$ data (blue open squares) and CMS NSD $pp$ data (red triangles) at the same centre-of-mass energy.
}
\label{figure:comparison-otherexpt}
\end{center}
\end{figure}

The \Nch\ distribution as a function of \pT\ 
in the kinematic range \pT~$>$ 500~MeV and $|\eta| <$~2.5 
is shown in Fig.~\ref{figure:comparison-otherexpt}.
The CMS~\cite{cmsminbias} results at the same centre-of-mass energy are superimposed.  The number of charged particles in the CMS data is consistently lower than the data presented in this paper.  This offset is expected from the CMS measurement definition of NSD events, where events with \nch~$=0$ enter the normalisation and the number of lower transverse momentum particles are reduced by the subtraction of the PYTHIA single diffractive component.  The UA1~\cite{Albajar:1989an} results, normalised by their associated cross section measurement, are also overlaid.  They are approximately 20\% higher than the present data.  A shift in this direction is expected from the double-arm scintillator trigger requirement used to collect the UA1 data, which rejected events with low charged-particle multiplicities.

%
%


To compare more directly the present data with results from CMS, the mean charged-particle density was calculated in the range $|\eta| <$~2.4 and a model dependent correction was applied to form an NSD particle density.  For the calculation of the NSD value the PYTHIA DW tune was selected due to its similarity with the tune used in the CMS analysis.    This generator set-up was used to produce a correction for the removal of the fraction of single diffractive events, the removal of electrons from \pizero\ Dalitz decays and the addition of non-single diffractive events with no charged particles within the kinematic range  \pT~$>500$~MeV and  $|\eta|<$~2.5.
The net effect of the correction is to reduce the charged-particle multiplicity.  
The resulting value 
1.240~$\pm$~0.040~(syst.)
is consistent with the CMS measurement of 
1.202~$\pm$~0.043~(syst.)
in the kinematic range of \pT~$>500$~MeV and  $|\eta|<$~2.4.

\section{Conclusions}

Charged-particle multiplicity measurements with the ATLAS detector using the first collisions delivered by the LHC during 2009 are presented.
Based on over three hundred thousand proton-proton inelastic interactions, 
the properties of events with at least one primary charged particle produced 
within the kinematic range $|\eta|<$ 2.5 and \pT~$>$~500~MeV
 were studied.  
The data were corrected with minimal model dependence to obtain inclusive distributions.
The charged-particle multiplicity per event and unit of pseudorapidity at $\eta =$~0 is measured to be 1.333$\pm0.003$(stat.)$\pm $0.040(syst.), which is 5--15\% higher than the Monte Carlo model predictions. 
The selected kinematic range and the precision of this analysis highlight clear differences between Monte Carlo models and the measured distributions.


\section{Acknowledgements}

We are greatly indebted to all CERN's departments and to the LHC project for their immense efforts not only in building the LHC, but also for their direct contributions to the construction and installation of the ATLAS detector and its infrastructure. 
All our congratulations go to the LHC operation team for the superb performance during this initial data-taking period.
We acknowledge equally warmly all our technical colleagues in the collaborating Institutions without whom the ATLAS detector could not have been built. Furthermore we are grateful to all the funding agencies which supported generously the construction and the commissioning of the ATLAS detector and also provided the computing infrastructure.

The ATLAS detector design and construction has taken about fifteen years, and our thoughts are with all our colleagues who sadly could not see its final realisation.

We acknowledge the support of ANPCyT, Argentina; Yerevan Physics Institute, Armenia; ARC and DEST, Australia; Bundesministerium f\"ur Wissenschaft und Forschung, Austria; National Academy of Sciences of Azerbaijan; State Committee on Science \& Technologies of the Republic of Belarus; CNPq and FINEP, Brazil; NSERC, NRC, and CFI, Canada; CERN;
CONICYT, Chile;
 NSFC, China; 
 COLCIENCIAS, Colombia;
 Ministry of Education, Youth and Sports of the Czech Republic, Ministry of Industry and Trade of the Czech Republic, and Committee for Collaboration of the Czech Republic with CERN; Danish Natural Science Research Council and the Lundbeck Foundation; European Commission, through the ARTEMIS Research Training Network; IN2P3-CNRS and Dapnia-CEA, France; Georgian Academy of Sciences; BMBF, HGF, DFG and MPG, Germany; Ministry of Education and Religion, through the EPEAEK program PYTHAGORAS II and GSRT, Greece; ISF, MINERVA, GIF, DIP, and Benoziyo Center, Israel; INFN, Italy; MEXT, Japan; CNRST, Morocco; FOM and NWO, Netherlands; The Research Council of Norway; Ministry of Science and Higher Education, Poland; GRICES and FCT, Portugal; Ministry of Education and Research, Romania; Ministry of Education and Science of the Russian Federation and State Atomic Energy Corporation ``Rosatom''; JINR; Ministry of Science, Serbia; Department of International Science and Technology Cooperation, Ministry of Education of the Slovak Republic; Slovenian Research Agency, Ministry of Higher Education, Science and Technology, Slovenia; Ministerio de Educaci\'{o}n y Ciencia, Spain; The Swedish Research Council, The Knut and Alice Wallenberg Foundation, Sweden; State Secretariat for Education and Science, Swiss National Science Foundation, and Cantons of Bern and Geneva, Switzerland; National Science Council, Taiwan; TAEK, Turkey; The Science and Technology Facilities Council and The Leverhulme Trust, United Kingdom; DOE and NSF, United States of America.

\clearpage
\newpage

\bibliographystyle{model1-num-names}
\bibliography{MinBiasFirstPub}

\clearpage
\newpage

\include{atlas_authlist}

\end{document}

%% file: atlas_authlist.tex
\begin{flushleft}
{\Large The ATLAS Collaboration}

\fontsize{8}{10}
\selectfont

\bigskip

G.~Aad$^{\rm 48}$,
E.~Abat$^{\rm 18a}$$^{,*}$,
B.~Abbott$^{\rm 110}$,
J.~Abdallah$^{\rm 11}$,
A.A.~Abdelalim$^{\rm 49}$,
A.~Abdesselam$^{\rm 117}$,
O.~Abdinov$^{\rm 10}$,
B.~Abi$^{\rm 111}$,
M.~Abolins$^{\rm 88}$,
H.~Abramowicz$^{\rm 151}$
% ...
\bigskip

% ...
$^{10}$ Institute of Physics, Azerbaijan Academy of Sciences, H. Javid Avenue 33, AZ 143 Baku, Azerbaijan\\
$^{11}$ Institut de F\'isica d'Altes Energies, IFAE, Edifici Cn, Universitat Aut\`onoma  de Barcelona,  ES - 08193 Bellaterra (Barcelona), Spain\\
% ...
$^{18}$ Bogazici University$^{(a)}$, Faculty of Sciences, Department of Physics, TR - 80815 Bebek-Istanbul; Dogus University$^{(b)}$, Faculty of Arts and Sciences, Department of Physics, 34722, Kadikoy, Istanbul; $^{(c)}$Gaziantep University, Faculty of Engineering, Department of Physics Engineering, 27310, Sehitkamil, Gaziantep, Turkey; Istanbul Technical University$^{(d)}$, Faculty of Arts and Sciences, Department of Physics, 34469, Maslak, Istanbul, Turkey\\
% ...
$^{48}$ Albert-Ludwigs-Universit\"{a}t, Fakult\"{a}t f\"{u}r Mathematik und Physik, Hermann-Herder Str. 3, D - 79104 Freiburg i.Br., Germany\\
$^{49}$ Universit\'e de Gen\`eve, Section de Physique, 24 rue Ernest Ansermet, CH - 1211 Geneve 4, Switzerland\\
% ...
$^{88}$ Michigan State University, Department of Physics and Astronomy, High Energy Physics Group, East Lansing, MI 48824-2320, United States of America\\
% ...
$^{110}$ University of Oklahoma, Homer L. Dodge Department of Physics and Astronomy, 440 West Brooks, Room 100, Norman, OK 73019-0225, United States of America\\
$^{111}$ Oklahoma State University, Department of Physics, 145 Physical Sciences Building, Stillwater, OK 74078-3072, United States of America\\
% ...
$^{117}$ Oxford University, Department of Physics, Denys Wilkinson Building, Keble Road, Oxford OX1 3RH, United Kingdom\\
% ...
$^{151}$ Tel Aviv University, Raymond and Beverly Sackler School of Physics and Astronomy, Ramat Aviv, IL - Tel Aviv 69978, Israel\\
% ...
$^{a}$ Present address FermiLab, USA\\
$^{b}$ Also at CPPM, Marseille, France.\\
$^{c}$ Also at TRIUMF, 4004 Wesbrook Mall, Vancouver, B.C. V6T 2A3, Canada\\
$^{d}$ Also at Faculty of Physics and Applied Computer Science of the AGH-University of Science and Technology, (FPACS, AGH-UST), al. Mickiewicza 30, PL-30059 Cracow, Poland\\
% ...
$^{*}$ Deceased\end{flushleft}

\fontsize{12}{15}
\selectfont